\documentclass{eage2020}

\usepackage{tikz}
\usepackage{float}
\usepackage{caption}
\usepackage{subcaption}

\begin{document}

\textbf{Comparing Strategies for Local FWI: FD Injection and Immersive Boundary Conditions}
\vspace{0.2cm}

\textbf{Fernanda Farias and Reynam Pestana}

\textbf{Abstract}

Conventional Full Waveform Inversion requires calculating the objective function to be
minimized and construction a gradient using the whole property model, when is often the case
where geoscientist are only interested in a local region. In this study, we use two strategies to
perform local FWI in time domain. One that disregards the interaction of the locally altered
model with the exterior (FD injection) and the second that can take these iterations into
account (Immersive Boundary Conditions). Numerical tests show the influence of whether or
not to set aside these interactions for different accuracy of the exterior model.

\clearpage

\section{Introduction}

The full waveform inversion (FWI) technique is often used to provide high resolution propriety model of the subsurface. For this purpose it is necessary to compute both the forward and the adjoint source extrapolation of the entire model, even when the geoscientist is usually interested in a small portion of the model. This can sound as a waste of calculation and, to save time and speed up calculations, attempts have been made to reproduce the wavefield inside a local domain as if it had been generated using the entire model. In this work, we compare the results of two different approaches to perform a localized FWI.

The first strategy uses the finite difference injection method proposed by \cite{robertsson2000efficient}, which basically stores the wavefields on the boundary of the local domain to inject then after alterations have occurred, whereas the surrounding model is kept unaltered. However, this method is unable to recover the high-order long-range interaction between the local and the surrounding model. Due to its lower cost and straightforward implementation, it has already been used
for FWI applications in 3D and with the elastic wave equation \citep{borisov2015efficient}.

The second and more computationally expensive approach, which we adopt here, makes use of the technique introduced by \cite{van2007exact} and extended by \cite{vasmel2013immersive} for two and three dimensions, the so-called immersive boundary condition (IBC), which is capable of mimic, under numerical precision, the wavefield in a domain which is immersed in a larger one. It requires a large number of Green's functions computations in the full domain to calculate the exact boundary conditions at every time step, and by doing so, this methodology can account for every interaction that the local domain may have with the exterior domain. Lately, IBC has been used in the frequency domain for salt boundary problems by \cite{willemsen2016numerically}, and it was also tested using a randomized singular value decomposition to compute the low-rank approximation of the Green's function in two and three dimension \citep{kumar2019enabling} .

To observe the effect of high-order long-range interaction for localized FWI applications, we compare the velocity model inverted using the two localized techniques mentioned above with the conventional inversion in the time domain using the vector acoustic equations. 

\section{Method and Theory}

\textit{FD Injection}

\begin{figure}[!htb]
\centering
\begin{subfigure}{.5\textwidth}
  \centering
  \includegraphics[width=.7\linewidth]{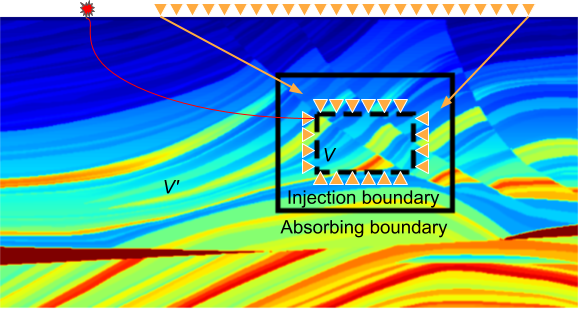}
  \caption{}
  \label{fig:sub1}
\end{subfigure}%
\begin{subfigure}{.5\textwidth}
  \centering
  \includegraphics[width=.7\linewidth]{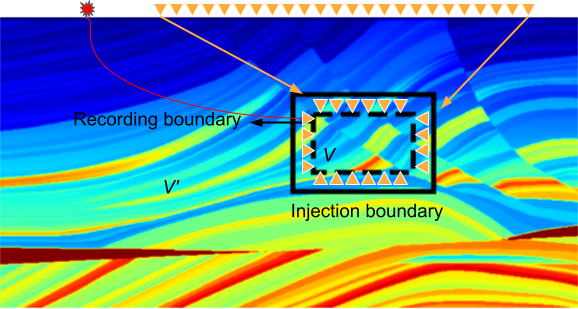}
  \caption{}
  \label{fig:sub2}
\end{subfigure}
\caption{Design of relevant surfaces for (a) the FD injection method. After modeling using the whole model, the wavefields are stored and later injected in the dashed rectangle (the solid rectangle acts as an absorbing boundary). (b) IBC concept: wavefield are extrapolated from the dashed surface to the solid one, where it is injected generating the exact boundary conditions.}
\label{Fig2}
\end{figure}

The finite-difference injection method used to model seismic data after local model alterations connects the solution of different regions by repositioning source and receiver wavefields on the surface of the local regions. This is done after one simulation using the whole model, in which both the pressure and the particle velocity wavefields are stored at the injection surface. After alterations in the local model, these wavefields will act as a new source distribution for the local simulations. To avoid undesired events, a larger surface should involve the injection surface as depicted in on the right side of Figure \ref{Fig2}(a), in a way that the localized simulations are carried out only inside the continuous line in Figure \ref{Fig2}(a). Both forward calculation and gradient constructions are carried out exclusively inside this local domain. The computation of the gradient requires redatuming, i.e., to replace receivers from the physical to the virtual position. This step is carried out for all four sides of the local domain. This can be done by extrapolating the observed wavefield from its recorded positions to the virtual receivers, assuming that the background velocity model used for calculating the Green function is sufficiently accurate. An alternative for redatuming for localised FWI is to use the method proposed by \cite{kohnke2019inversion}, who introduced an iterative least-squares problem linking the wavefield on the local domain surface to the observation data.

\textit{Immersive Boundary Conditions}

To overcome the inability of the FD injection method to account for the high-order interactions between the scattered wavefield and the unaltered exterior domain, the so-called immersive boundary condition calculates the boundary conditions at the limit of the local model "on-the-fly" so that nothing is neglected. To do that, an auxiliary transparent surface within the local model is defined (see Figure \ref{Fig2}(b)), after that the wavefield that is recorded in that auxiliary surface is propagated to the injection surface using a time-recursive discrete version of the   Kirchoff-Helmholtz integral:

\begin{equation}
    p(\textbf{x},t) = \int_S [p(\textbf{x}^S,t) * G^{p,f}(\textbf{x},\textbf{x}^S,t) + \textbf{v}(\textbf{x}^S,t) * G^{p,q}(\textbf{x},\textbf{x}^S,t)] \cdot \textbf{n} dS
    \label{eq:ibc}
\end{equation}

The wavefield reconstructed at the $S_{inj}$ surface is updated and injected at every time step and it carries every interaction with the extended model through the Green's functions presented in Equation \ref{eq:ibc}. The Green's functions are calculated at every point on the record surface and stored at the injection surface, although \cite{broggini2018sensitivity} made a sensitivity study concerning the spatial subsampling on the recording surface. In either way, the convolutions present in Equation \ref{eq:ibc} can be quite time consuming depending on the model parameters. For a detailed description of the implementation and theory of the IBC methodology we refer to \cite{broggini2017immersive}.

\textit{Full Waveform Inversion using Vector Acoustic Equations}

Although FWI using vector acoustic equations are still less usual than the conventional second order acoustic wave equation, a few studies show the applications of this technique to synthetic \citep{akrami2017algorithm} and field data \citep{hwang20192d}, attempting to explore the benefits of using direction data into the inversion. Here, we use the pressure field to fit the data, using the Equation \ref{of}, and the gradient expression (Equation \ref{grad}) to construct it locally as proposed here.



\begin{equation}
    J(m) = \frac{1}{2} \sum_{s,r} \int_T ||\textbf{W}_r [\textbf{u}_q(\textbf{x}_s,\textbf{x}_r,t;m) - \textbf{d}_q(\textbf{x}_s,\textbf{x}_r,t)] ||^2 dt
    \label{of}
\end{equation}

\begin{equation}
    \frac{\partial J}{\partial m} (m) = \sum_{s} \int_T \frac{\partial p}{\partial t} \phi \textsubscript{q} dt
    \label{grad}
\end{equation}

\section{Examples}

\textit{Horizontal reflector model}

\begin{figure}[!htb]
  \centering
  \begin{subfigure}[b]{0.24\textwidth}
    \centering
    \includegraphics[width=4cm, height=3cm]{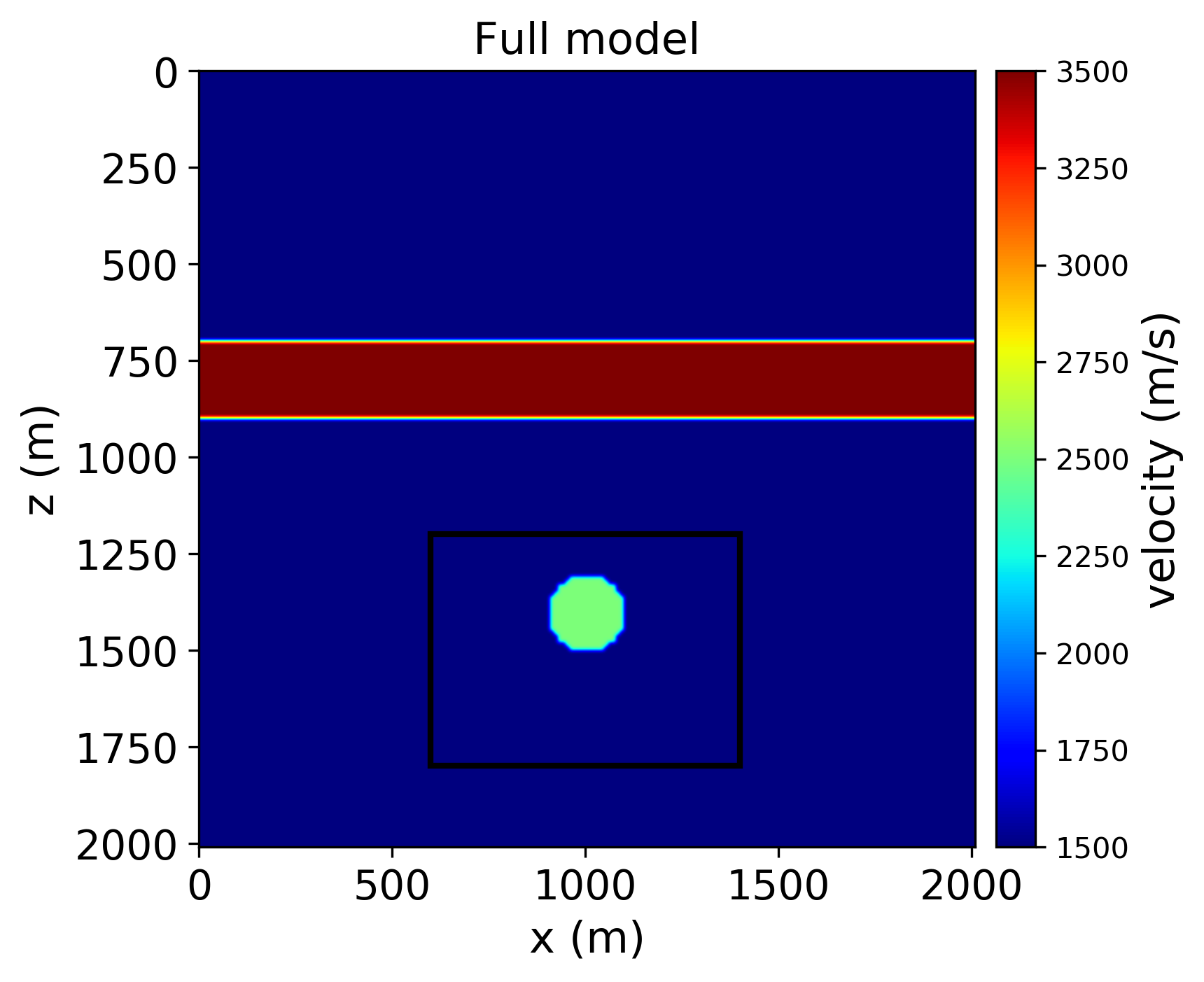}
    \caption{}
    \label{Fig3.1}
  \end{subfigure}
  \hspace{.25cm}
  \begin{subfigure}[b]{0.3\textwidth}
    \centering
    \includegraphics[width=4cm, height=3cm]{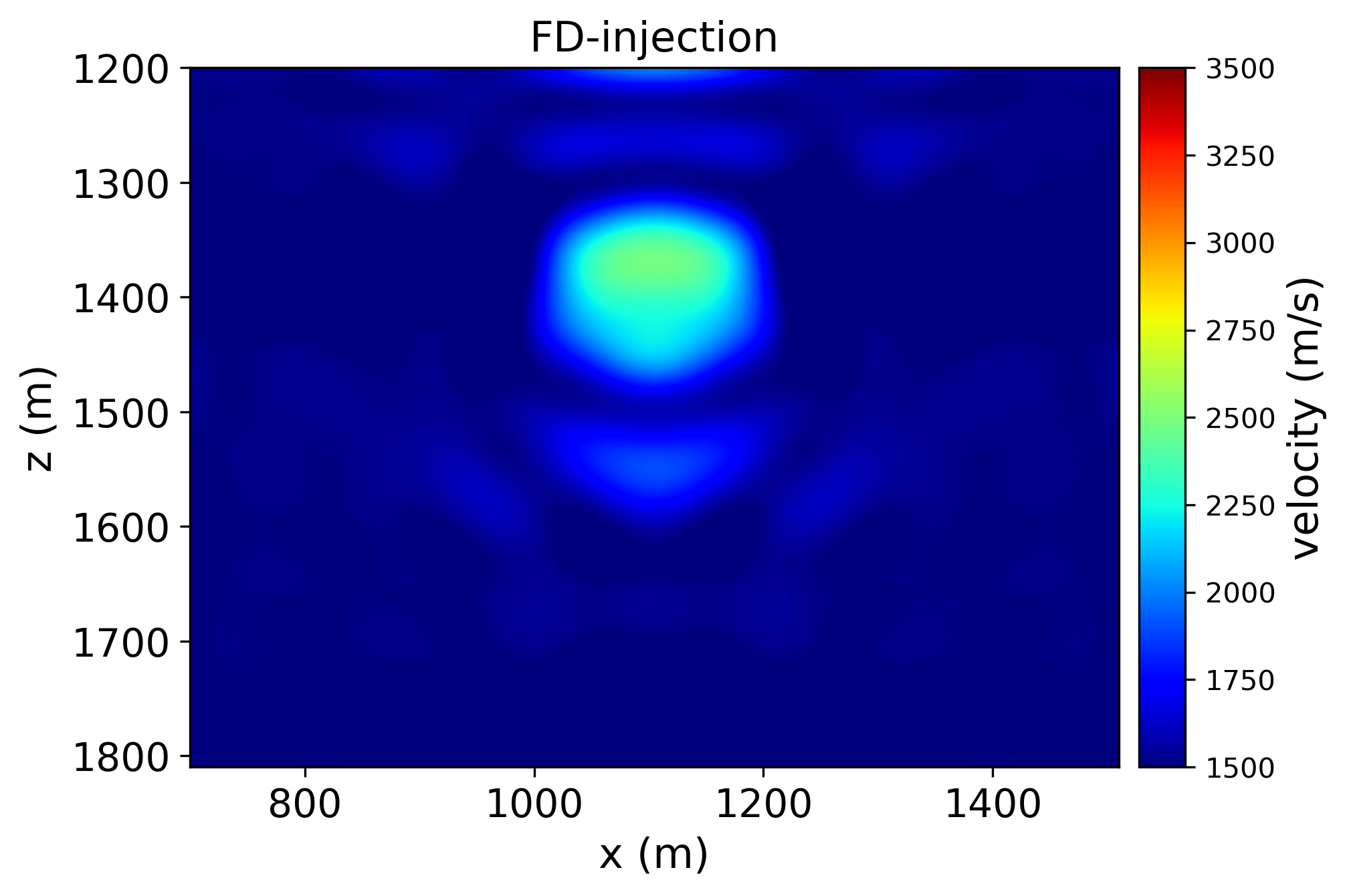}
    \caption{}
    \label{Fig3.2}
  \end{subfigure}
  \begin{subfigure}[b]{0.3\textwidth}
    \centering
    \includegraphics[width=4cm, height=3cm]{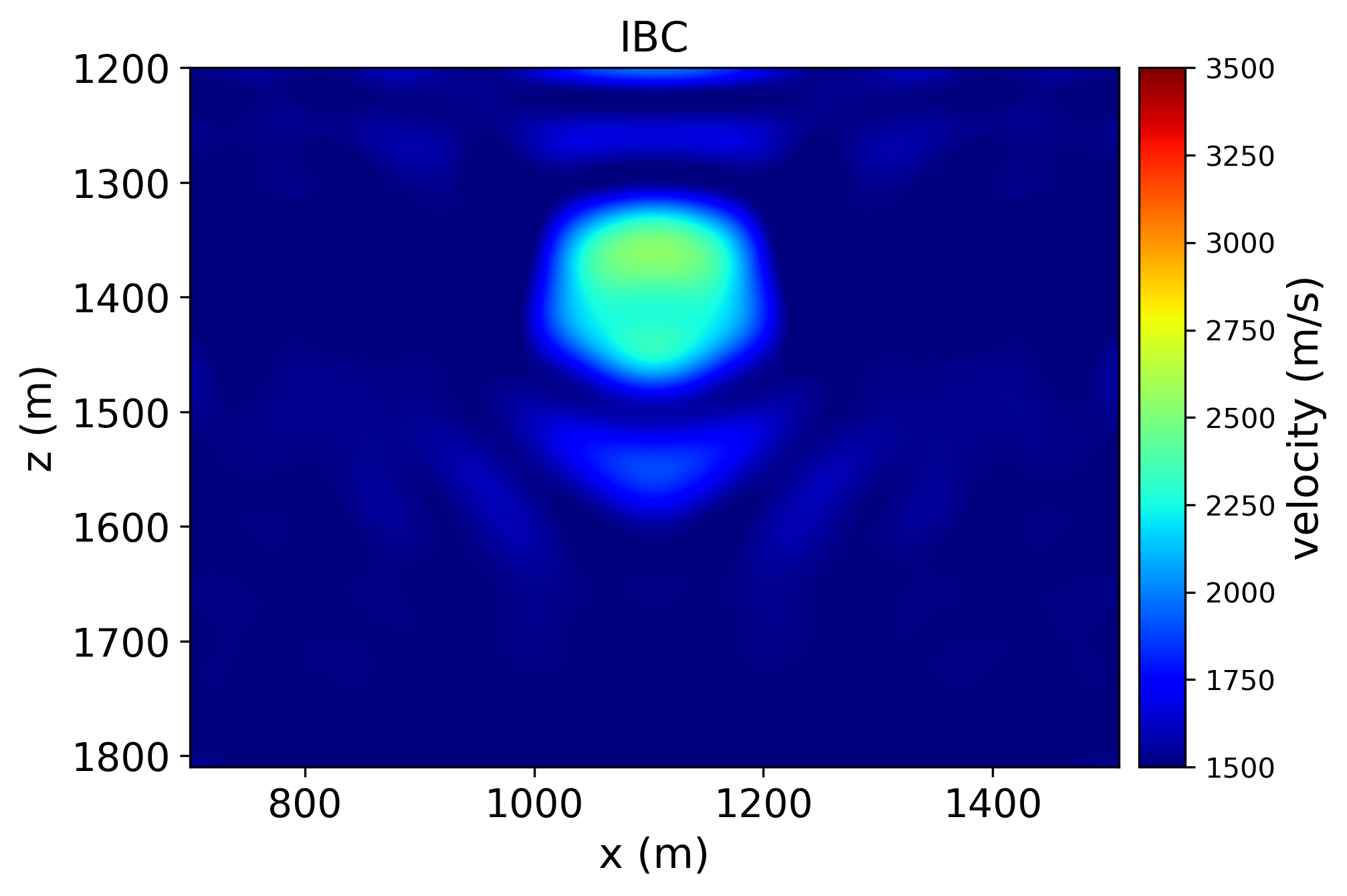}
    \caption{}
    \label{Fig3.3}
  \end{subfigure}
  \caption{(a) True velocity model. Final velocity model for the FD injection FWI (b) and IBC FWI (c) after 20 iterations using the conjugate gradient method for the horizontal layer model.}
  \label{Fig3}
\end{figure}

For the simple horizontal reflector model, displayed in Figure \ref{Fig3}(a), the background was exact the same as the true model except for the round anomaly below the reflector. This particular case, is a good scenario for IBC methodology. Since the background velocity model is the same as the true model outside the localized inversion, IBC can perfectly predict the interaction between the local model and the outside, which is normally not true. In this example, the density model is considered constant. Each source is placed at a depth of 10 m  and regularly spaced at 300 m interval. The peak frequency is set to 5 Hz.

\begin{figure}[!htb]
  \centering
  \begin{subfigure}[b]{0.30\textwidth}
    \centering
    \includegraphics[width=\textwidth]{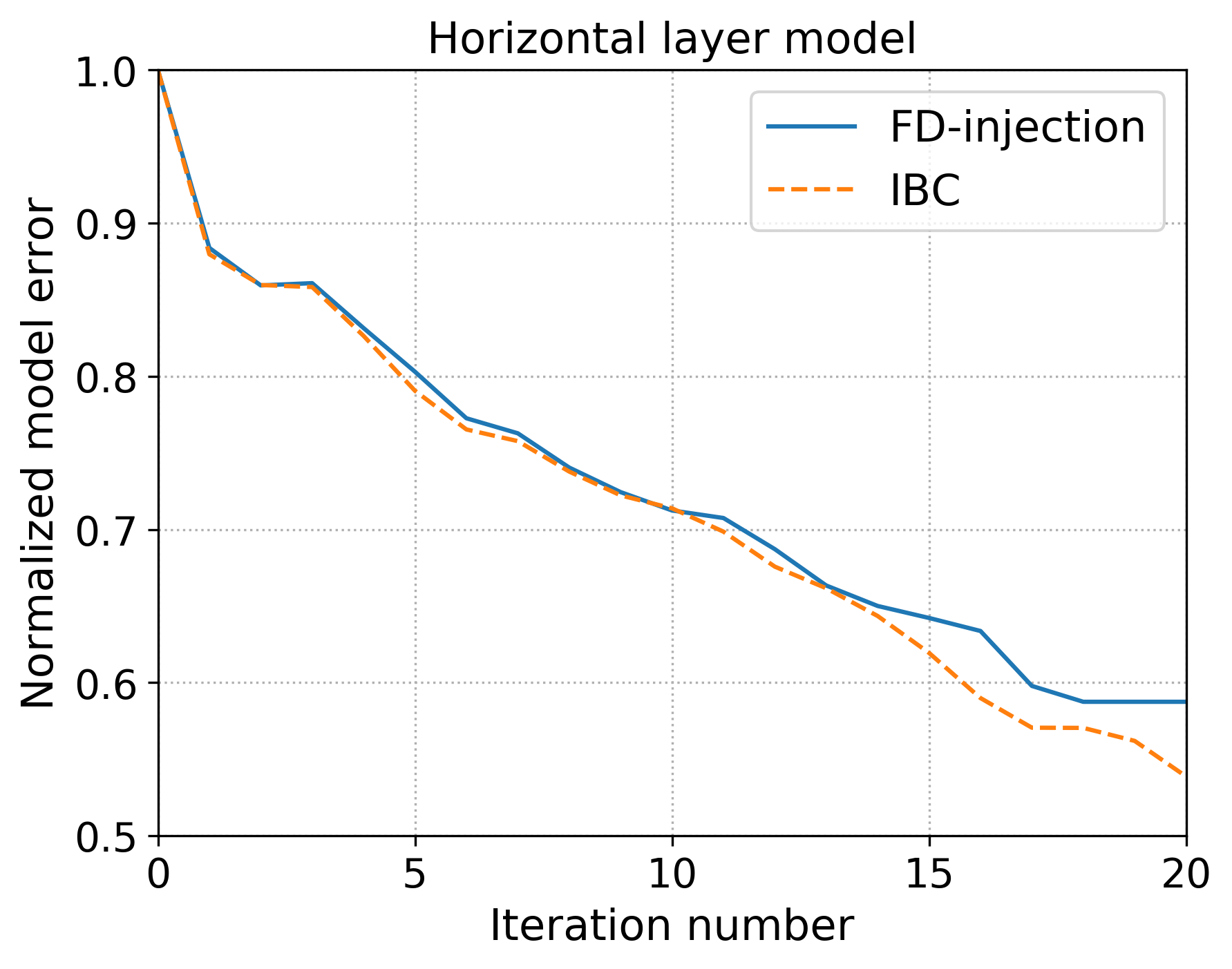}
    \caption{}
    \label{Fig4.1}
  \end{subfigure}
  \hspace{.15cm}
  \begin{subfigure}[b]{0.31\textwidth}
    \centering
    \includegraphics[width=\textwidth]{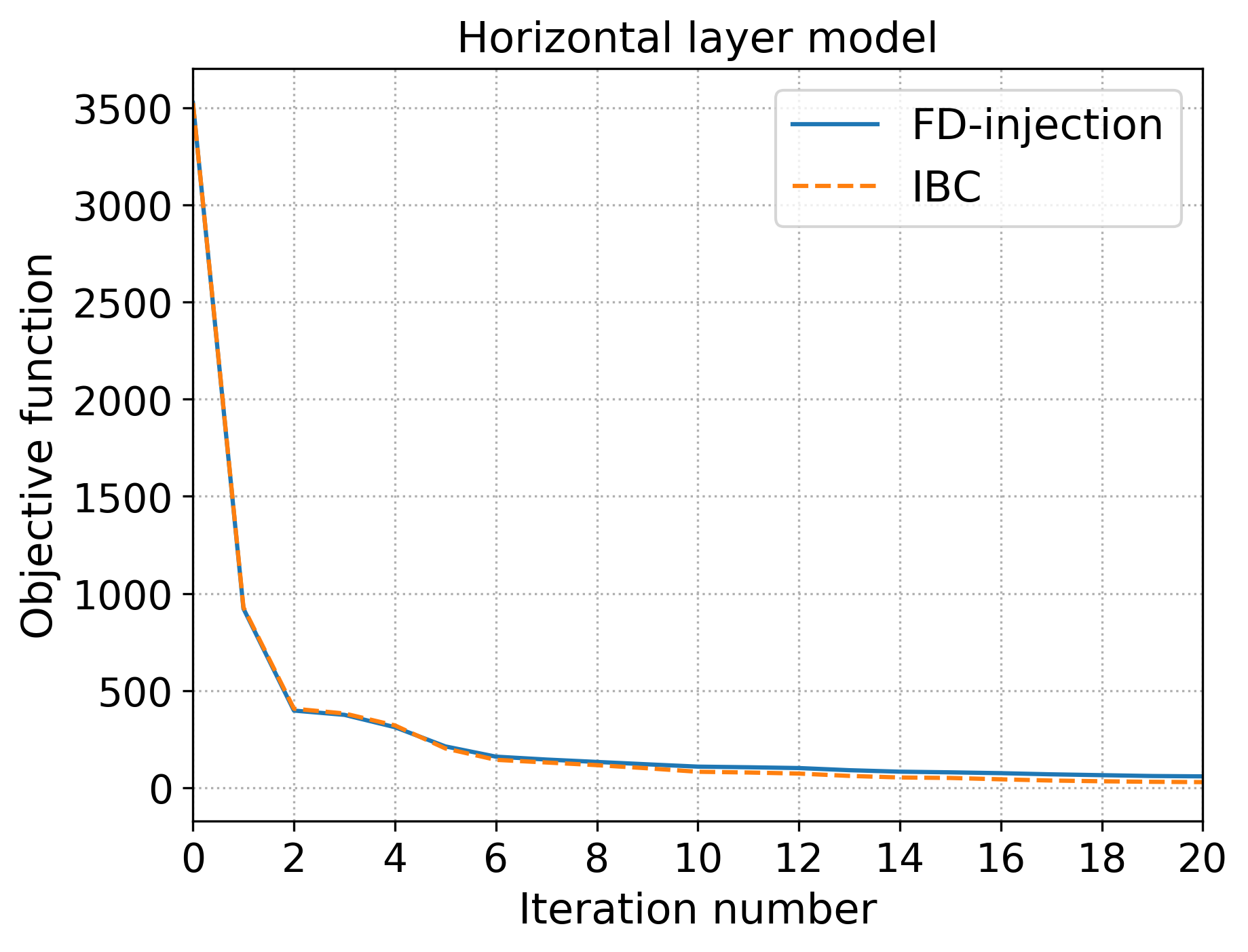}
    \caption{}
    \label{Fig4.2}
  \end{subfigure}
  \caption{Normalized model error (a) and objective function (b) as a function of the number of iterations of the conjugate gradient method for the horizontal layer model.}
  \label{Fig4}
\end{figure}

\textit{Marmousi model}

To investigate how both methods behave in case the model outside the local domain is inaccurate, which is usually true, we perform local FWI for the Marmousi model considering as local domain the region inside the rectangle in Figure \ref{Fig2}(b). As for the horizontal layer model, the source is placed at a depth of 10 m  and regularly spaced at 300 m interval, with peak frequency of 5 Hz. 

\begin{figure}[!htb]
  \centering
  \begin{subfigure}[b]{0.3\textwidth}
    \centering
    \includegraphics[width=\textwidth]{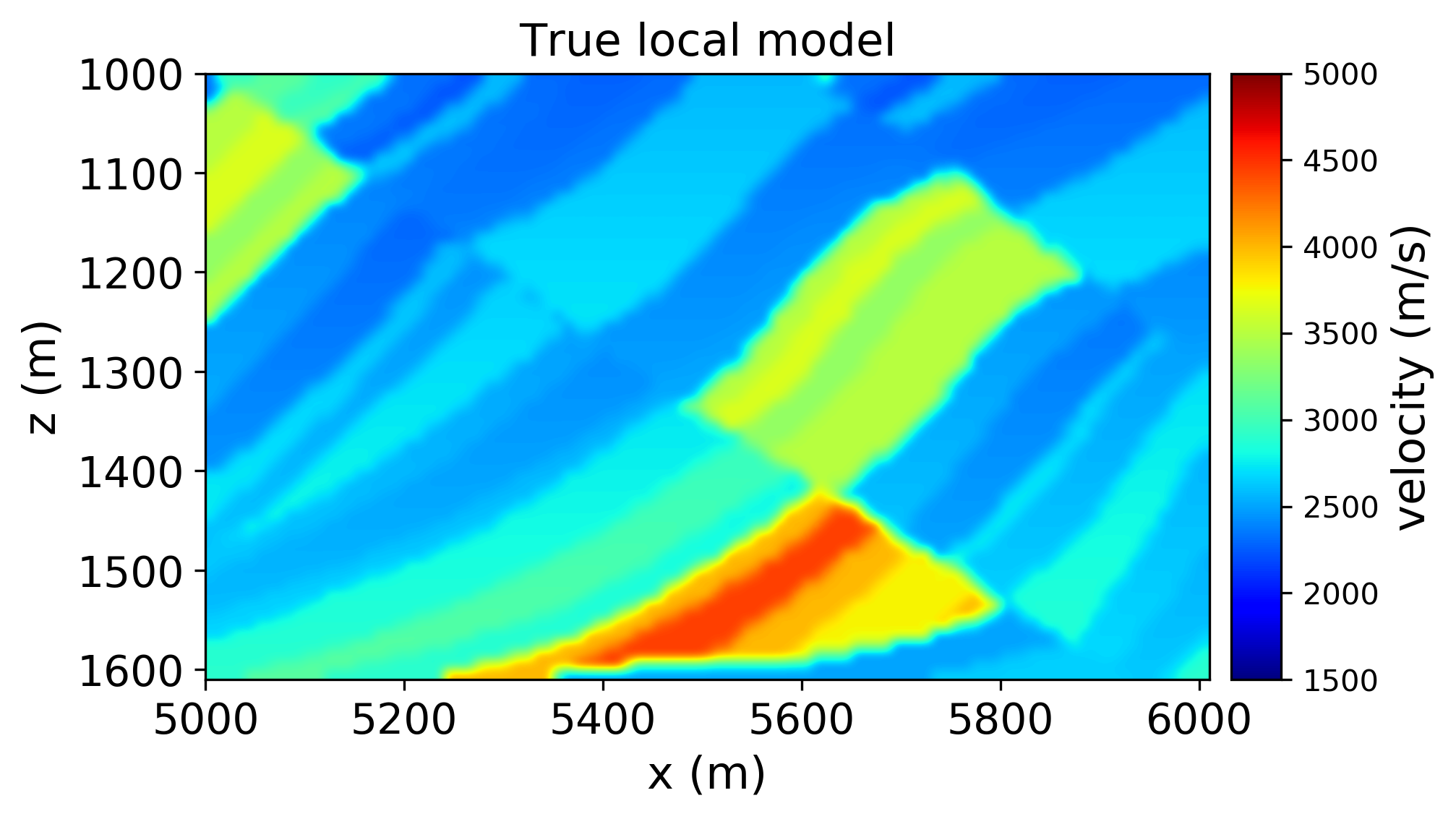}
    \caption{}
    \label{Fig7.1}
  \end{subfigure}
  \hspace{.15cm}
  \begin{subfigure}[b]{0.3\textwidth}
    \centering
    \includegraphics[width=\textwidth]{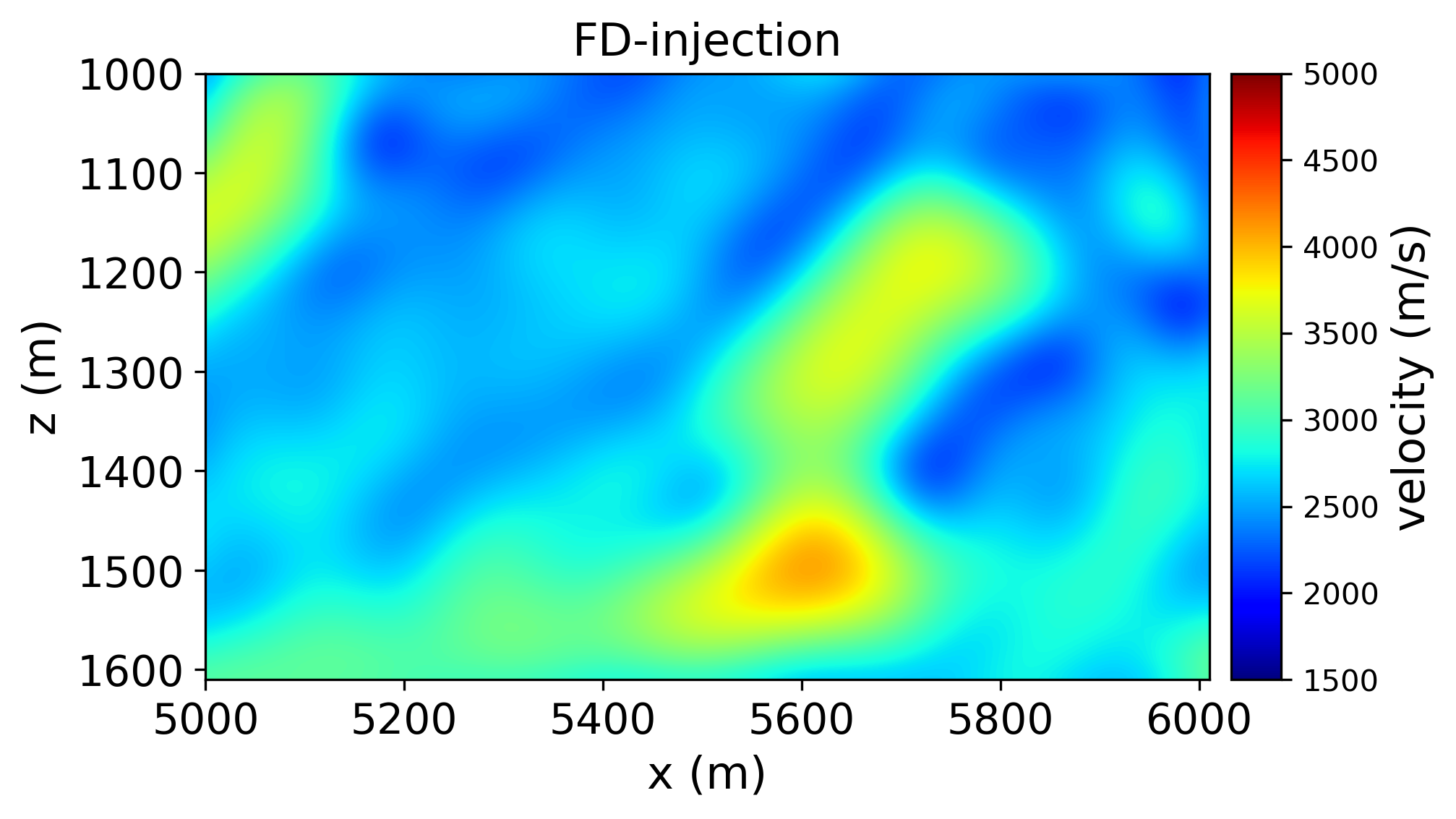}
    \caption{}
    \label{Fig7.2}
  \end{subfigure}
  \hspace{.15cm}
  \begin{subfigure}[b]{0.3\textwidth}
    \centering
    \includegraphics[width=\textwidth]{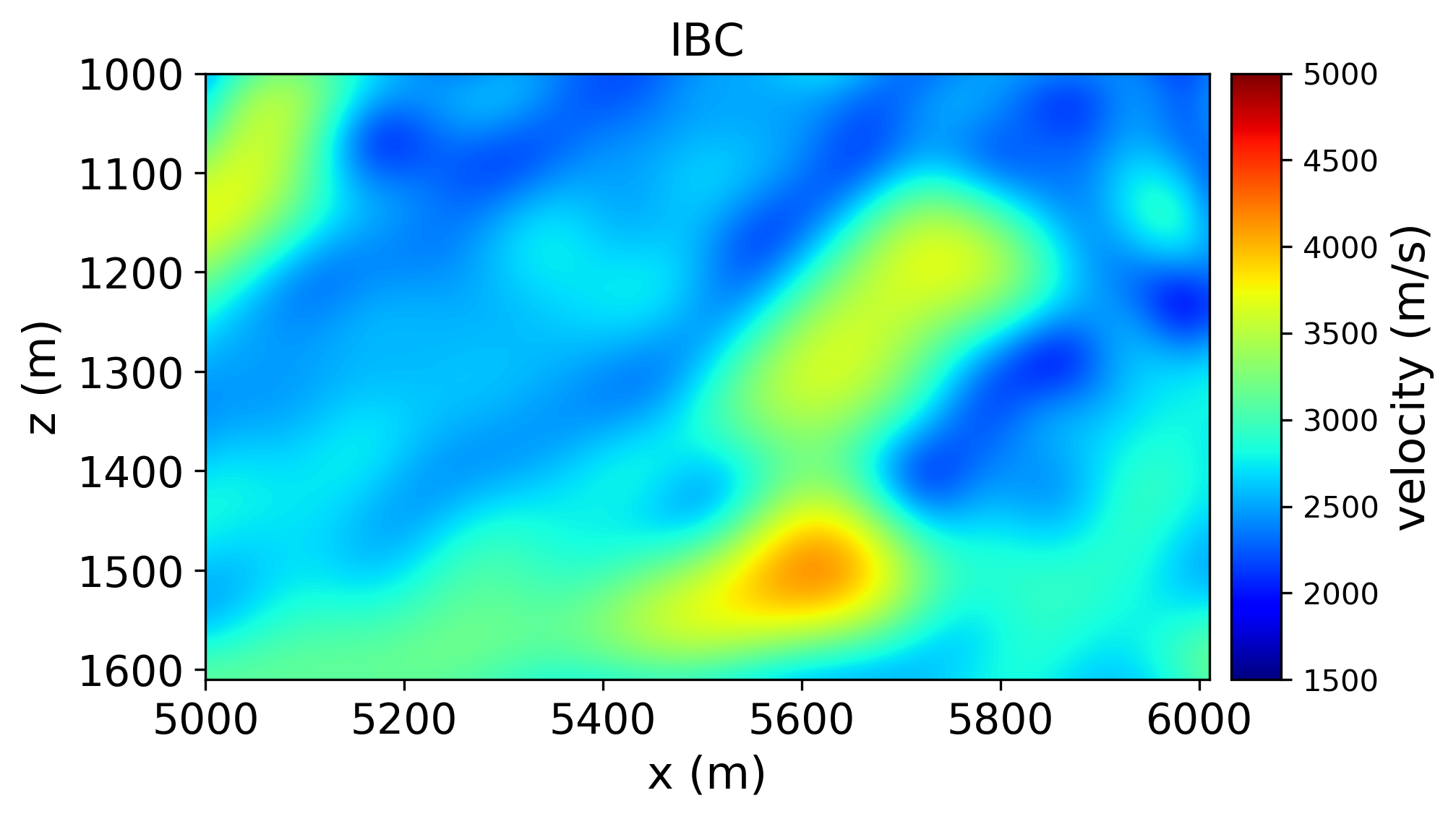}
    \caption{}
    \label{Fig7.3}
  \end{subfigure}
  \caption{(a) True local velocity model. Final velocity model for the FD injection FWI (a) and IBC (b), after 20 iterations using the conjugate gradient method for the marmousi model.}
  \label{Fig7}
\end{figure}

\begin{figure}[!htb]
  \centering
  \begin{subfigure}[b]{0.3\textwidth}
    \centering
    \includegraphics[width=\textwidth]{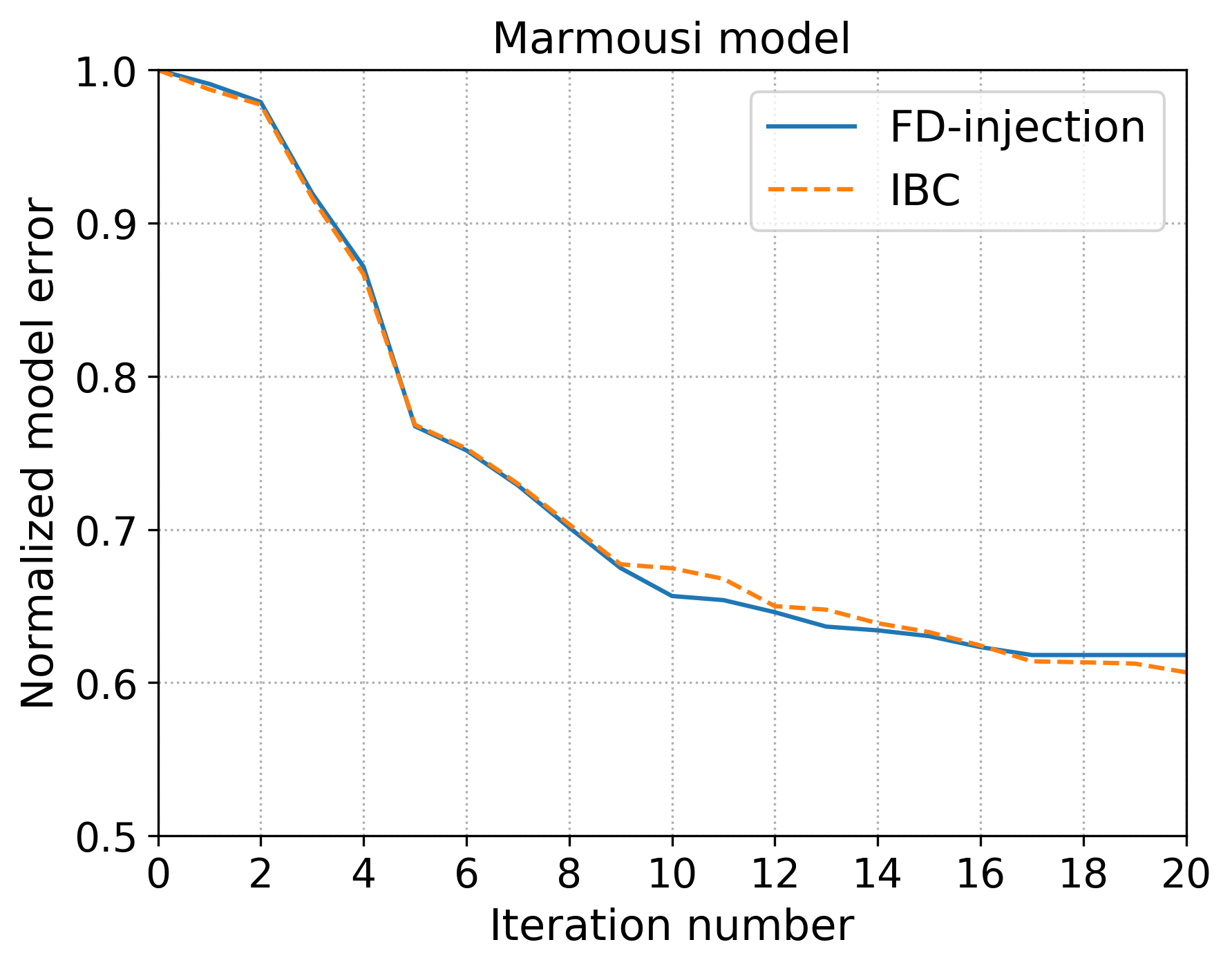}
    \caption{}
    \label{Fig8.1}
  \end{subfigure}
  \hspace{.15cm}
  \begin{subfigure}[b]{0.31\textwidth}
    \centering
    \includegraphics[width=\textwidth]{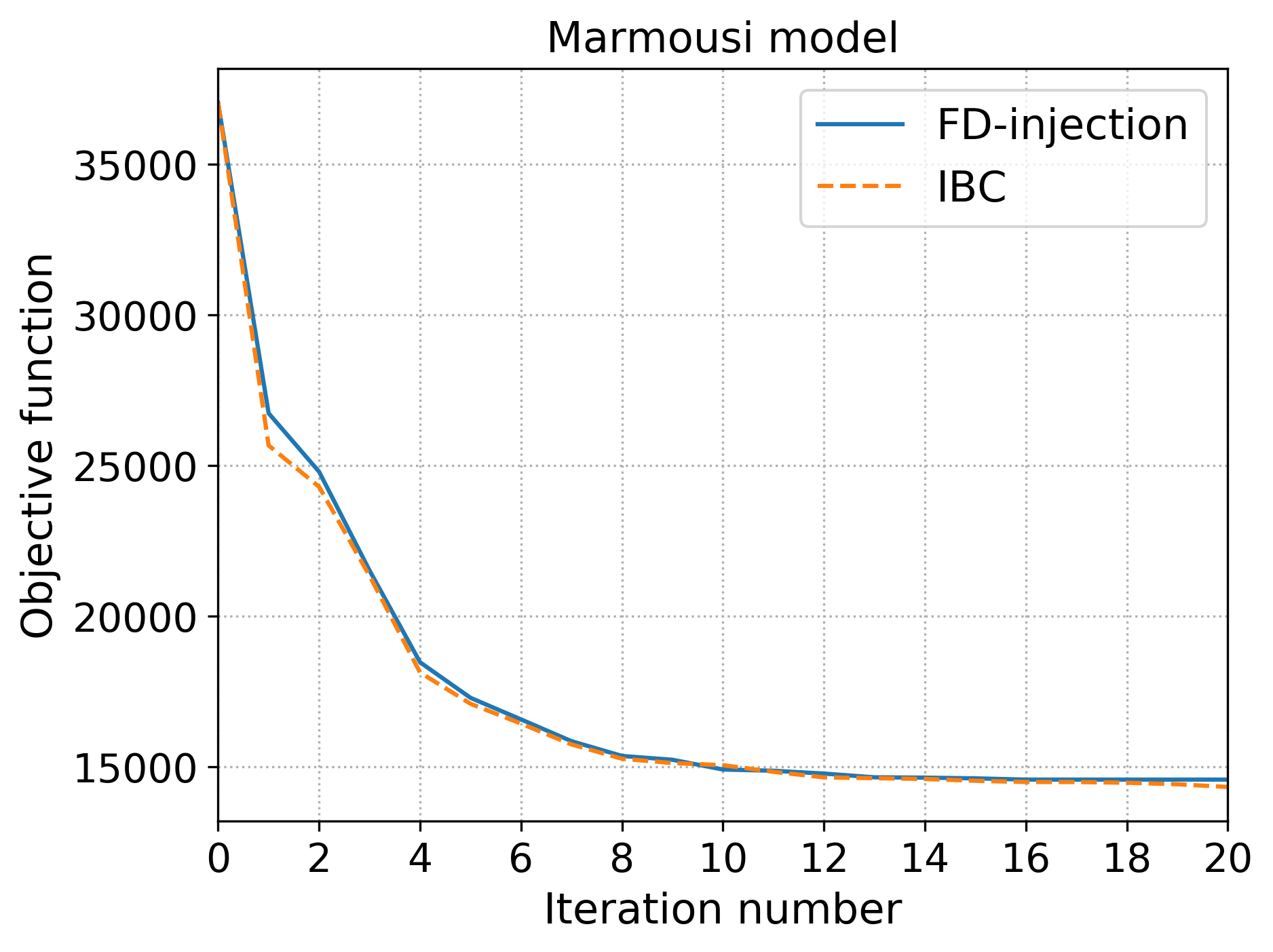}
    \caption{}
    \label{Fig8.2}
  \end{subfigure}
  \vspace{.001cm}
  \caption{Normalized model error (a) and objective function (b) as a function of the number of iterations of the conjugate gradient method for the marmousi model.}
  \label{Fig8}
\end{figure}

For the two models considered, the local FWI using IBC shows lower values for the normalized error model and objective function value at the end of 20 iterations than the one using the FD injection. Although, for the parameters considered here, inverted velocities are reasonable recovered using both methodologies.

\vspace{-.1cm}
\section{Conclusions}

Numerical tests indicate that when the overburden is the same as the one where the observed data was acquired, a convenient situation for time-lapse inversion since physical properties are likely to change in the reservoir area, FWI based on FD injection and IBC provide a reasonable inverted velocity model, although IBC shows lower model error and objective function values. In the case where we use as background velocity model a smoothed version of the true model, the IBC still gives better results but not very different from the one using the FD injection methodology.

\section{Acknowledgements}

We would like to thank Filippo Broggini for very helpful discussions. We acknowledge Alan Souza from PETROBRAS and Oscar Mojica from SENAI CIMATEC for discussions and help with numerical tests.

\bibliography{refs}

%
%

\end{document}